\documentclass[nofootinbib,showpacs,aps,10pt,reprint]{revtex4-1}

\usepackage{amssymb}
\usepackage{amsmath}
\usepackage{graphics}
\usepackage{epsfig}
\usepackage[dvips]{color}
\usepackage{bm}
\usepackage{epsfig,graphics,graphicx}

\newcommand{\be}{\begin{equation}}
\newcommand{\ee}{\end{equation}}
\newcommand{\bea}{\begin{eqnarray}}
\newcommand{\eea}{\end{eqnarray}}

\newcommand{\Eq}[1]{{Eq.~({\ref{#1}})}}

\newcommand{\sumint}{\sum\!\!\!\!\!\!\!\!\int}
\newcommand{\slsh}[1]{\not \! #1}

\begin{document}

\title{Warm inflation in the presence of magnetic fields}


\author{Gabriella Piccinelli$^*$, \' Angel S\' anchez$^\dagger$, Alejandro Ayala$^{**\ddagger}$, Ana Julia Mizher$^{**}$}{\affiliation{$^*$Centro Tecnol\'ogico, FES Arag\' on, Universidad Nacional Aut\' onoma de M\' exico, Avenida Rancho Seco S/N, Bosques de Arag\' on, Nezahualc\' oyotl, Estado de M\' exico 57130, M\' exico.\\
$^\dagger$Facultad de Ciencias, Universidad Nacional Aut\' onoma de M\' exico, Apartado Postal 50-542, M\'exico Distrito Federal 04510, M\'exico.\\
$^{**}$Instituto de Ciencias Nucleares, Universidad Nacional Aut\' onoma de M\' exico, Apartado Postal 70-543, M\' exico Distrito Federal 04510, M\' exico.\\
$^\ddagger$Centre for Theoretical and Mathematical Physics, and Department of Physics,
University of Cape Town, Rondebosch 7700, South Africa.}

\begin{abstract}

We study the effects of primordial magnetic fields on the inflationary potential in the context of a warm inflation scenario. The model, based on global supersymmetry with a new-inflation-type potential and a coupling between the inflaton and a heavy intermediate superfield, is already known to preserve the flatness required for slow-roll conditions even after including thermal contributions. Here we show that the magnetic field makes the potential even flatter, retarding the transition and rendering it smoother.

\end{abstract}


\maketitle

\section{Introduction}

Magnetic fields are ubiquitous in the universe. They have been observed up to galaxy clusters \cite{clusters} and superclusters \cite{superclusters} (for comprehensive reviews see Refs. \cite{OBS_review}) and there is indirect evidence, from gamma-ray observations of blazars \cite{Blazars}, of a pervasive intergalactic magnetic field, with a lower bound of $10^{-16} - 10^{-15}$ Gauss. Despite their widespread presence, their origin is currently unknown. There are two basic possibilities for their generation: they could be either primordial or produced during processes associated to large scale structure formation. The growing observational evidence for the presence of magnetic fields at all astrophysical scales strengthens the idea of the primordial origin of cosmic magnetism. This possibility has in turn two related consequences for the cosmological model: on the one hand, the mechanisms for their generation, development and amplification have to be found and on the other, it is necessary to consider their contribution, together with temperature corrections, into the scenario of the very early universe evolution.

A series of mechanisms for the early generation of magnetic fields have been proposed \cite{Turner-Widrow,Dolgov,EWPT,QCD} (see also Ref. \cite{Genesis_review} for a review). However, none of these mechanisms is problem free. The difficulty is to obtain both the required scale and amplitude to match the presently observed fields. Some mechanisms have also addressed the generation of magnetic fields during inflation. One of these relies on the fact that large-scale cosmological magnetic fields can be created by the same mechanism that generates density fluctuations, i.e., quantum fluctuations in the Maxwell field that are excited inside the horizon and are expected to freeze-out as classical electromagnetic waves once they cross the Hubble radius. These initially static electric and magnetic fields can subsequently lead to current supported magnetic fields, once the excited modes reenter the horizon.  Nevertheless, the problem is to tie these seeds to present observations since magnetic fluctuations that survive a period of de Sitter expansion are typically too weak to match the present observations, as long as magnetic fields decay adiabatically with the universe expansion. To avoid this huge suppression, either a mechanism that breaks conformal invariance of electromagnetism must be introduced \cite{Turner-Widrow}, \cite{Ratra—Bamba} (see however \cite{Giovannini-Shap}), or we have to rely on the fact that the curvature of the background space can modify this adiabatic decay law \cite{Tsagas}. Another possible mechanism is to consider that non-Abelian gauge theories may have a ferromagnet-like vacuum (Savvidy vacuum), with a non-zero magnetic field, even at high temperatures \cite{Savvidy,Berera1999}. The formation of this non-trivial vacuum state at GUT scales can give rise to a Maxwell magnetic field imprinted on the comoving plasma.  An obstacle to inflationary magnetogenesis is the so-called back reaction problem \cite{back_react}, that consists on the observation that the generation of magnetic fields during inflation increases the energy density of the electromagnetic field, which can eventually dominate over the inflaton energy.

Whether or not these primordial magnetic fields (PMF) survive up to the present epoch and are able to match the observed cosmic fields, there are good chances that they were present during the early universe, where phase transitions also provided suitable conditions for their generation, such as charge separation, turbulence and departure from equilibrium. PMF directly interact with baryons, modifying these particles' evolution and velocity, and indirectly influence photons, through the tight coupling between baryons and photons at that epoch. Cold dark matter is also indirectly affected, through gravitational interaction. In order to gain some insight on the features and strength of PMF, one can resort to cosmic observational events to provide clues for the building of a cosmological magnetic model.

The effects of PMF on cosmological phase transitions have been widely studied. These phase transitions include the electroweak phase transition \cite{ellos,Querts,nosotros}, with particular emphasis on the baryogenesis process (See \cite{Piccinelli04} for a review; see also \cite{Comelli99}), or the supersymmetry phase transition \cite{Salam-Aky-Roy}. The studies include also the effects of PMF on the cosmic background radiation (see e.g. \cite{Yamazaki12} and references therein), on the nucleosyhtesis process (a detailed review can be found in \cite{Dario_review}), on structure formation \cite{Yamazaki06} and on the primordial gravitational waves spectrum \cite{Caprini}.

Assuming that the PMF power spectrum $P_B (k)$ depends on $k$ as a simple power law function on large scales, PMF are fully described by two parameters:  the spectral index, $n_B$ (an important parameter for the discrimination between models of magnetogenesis) and the root-mean-square of the field smoothed over length scale $\lambda$, $B_\lambda$. Alternatively, in some works, bounds on the total magnetic field energy density are found, and not on the smoothed amplitude of the magnetic field.  Limits obtained at different events typically involve different coherence scales, which are related to the Hubble horizon size at that epoch. The reported strengths are usually scaled to present values ($B_0$), assuming adiabatic evolution.

From nucleosynthesis, an upper bound of $B_0 \leq 3 \times 10^{-7}$ G, at length scales of the order of the Hubble horizon size at BBN time (which today corresponds approximately to 100 pc) \cite{Dario95} or an updated value $\langle B_0 \rangle \leq 1.5 \times 10^{-6}$ G (related to the contribution to the local field amplitude $B$ from all wavelengths)  \cite{Kawasaki12} can be found. From the large-scale structure formation process, the imprints of PMF can be searched for through the thermal-SZ effect, leading to the bound $B_0 \sim 10^{-8}$ G (see e.g. \cite{Tashiro12} and references therein), the Lyman-alpha forest: $B_0  \sim 10^{-9}$ G,  at scales 1 Mpc for a range of nearly scale invariant models, corresponding to magnetic field power spectrum index $n \simeq -3$  \cite{Pandey12}, or the matter power spectrum, leading to the bound of $B_0 \sim 1.5 - 4.5 \times 10^{-9}$ G and $n_B \in [-3, -1.5]$, considering the total magnetic field energy density \cite{Kahniashvili13}.

A lot of work has been done in the field of the observational
constraints considering different aspects of the interaction of
primordial magnetic fields with the CMB, as well as different features
and scales of these cosmic fields (see for example \cite{Giovannini06}
and references therein). Constraints have been derived using the CMB
temperature and polarization power spectra \cite{Planck13-XVI,Power-spectra}, Faraday rotation \cite{Faraday} and studying its non-Gaussian correlations, considering the bispectrum \cite{NG-bispectrum}, as well as the trispectrum \cite{NG-trispectrum}. The upper bounds that are established are between a few and tenth of nano Gauss. Let us consider, in particular, the last bounds obtained from Planck data. The constraints with the Planck+WP likelihood, where WP stands for WMAP9 large-scale polarization likelihood, are $B_{1 Mpc} < 4.1$ nG, with a preference for negative spectral indices at the $95\%$ confidence level. These limits are improved using Planck+WP+highL to $B_{1Mpc} < 3.4$ nG, where highL means that data from ACT (Atacama Cosmology Telescope)  and SPT (South Pole Telescope) at high angular scales are used \cite{Planck13-XVI}. These new constraints are consistent with, and slightly tighter, than previous limits from CMB.

In view of these constraints, in this work we consider the possibility of having PMF with magnitudes of the order of nano Gauss, scaled to present days.  With these values, the hierarchy of scales in the problem is established as $ eB < m^2 \ll T^2$, with $m$ a characteristic mass scale and $T$, the temperature.

Inflation has been considered both as a process that constrains as well as one able to generate PMF, whereas, as far as we know, the implications on the cosmic inflation potential of these PMF have not yet been explored.  In particular, given that the warm inflation model aims to fully include the effect of all the interactions on the inflaton dynamics, it is important to consider these PMF at the inflation epoch. In this work we study the effects of PMF on the warm inflationary potential.
	
The paper is organized as follows: In Sec. II, we recall the basic features of the warm inflation scenario and recall the calculation involving purely thermal
effects~\cite{HM}. In Sec. III, we introduce the formalism that allows to include the thermo-magnetic contributions to the self-energies of the heavy sector and through them the modification to the vacuum energy. We show that the magnetic field contributes to the flattening of the inflation potential and thus preserves the conditions for slow roll. Finally we summarize and conclude in Sec. IV. We leave for the appendices the explicit calculations of the thermo-magnetic corrections to the heavy sector masses and effective potential in the presence of a magnetic background.

\section{The model}

\subsection{Warm inflation}

Early models of inflation -dubbed super-cooled models (see e.g. \cite{Olive} for a review)- assumed very little interaction of the inflaton with all other fields until the reheating process, at the end of inflation. With the proposal of warm inflation \cite{Berera, Moss}, this picture changed: the inflaton is now assumed to interact with other fields, both during the inflationary expansion as well as at reheating, in a continuous and more natural way. It is a model where (near) thermal equilibrium conditions are maintained during the inflationary expansion, with no need for very flat potentials, nor for a tiny coupling constant. The model does require a dissipative component $\Gamma$ of sizable strength as compared to the expansion rate of the universe. This is opposed to the standard inflationary scenario where the damping term comes only from the universe's expansion. This additional dissipation is responsible for producing radiation since during an exponential expansion, the dissipation is dissolved very quickly and a source of radiation is needed. In this way, the equation of motion for the inflaton $\phi$ becomes

\begin{equation}
\ddot\phi+(3H+\Gamma)\dot\phi+V_{T,\phi}=0,\label{wip}
\end{equation}
where $H$ is the Hubble parameter,  and $V_{T, \phi}$ is the derivative with respect to $\phi$ of the inflaton effective potential (usually taken as the finite temperature one-loop Coleman-Weinberg potential). Warm inflation requires $\Gamma>3H$.

On the observational side, Planck's results establish a series of constraints on the abundant family of inflationary potentials proposed up to now \cite{Planck-infl-NG}. Given that the lack of detection of primordial non-gaussianites is a robust experimental result, stringent bounds on a series of inflationary scenarios can be established, among these, warm inflation. It follows that the strongly-dissipative regime required for this scenario is constrained, nevertheless this remains still viable (see however \cite{Hall-Peiris} and references therein, for a possible overproduction of gravitinos). It is interesting to note that not only the potentials typically employed in warm inflation models (new, natural, hybrid-type inflationary potentials) survive the severe Planck's analysis, but also that some potentials that are essentially ruled out by Planck in the context of cold inflation, are completely in agreement with observations when the inflation evolves in a thermal bath \cite{being-warm, Mar1, Enqvist}.

Since in warm inflation radiation is produced during the whole epoch of inflation, light fields, associated to this radiation, must be present in the Lagrangian. In these models, the inflaton interacts all the time with other fields, but its direct interaction with the light fields brings up some inconsistencies. Since the inflaton has a large expectation value, fields that interact directly with it acquire large masses. This fact is inconsistent with the radiation-like nature of such fields. Alternatively, one could limit the value of the coupling between the inflation and the light fields. However this would require an extremely low upper bound for the coupling, which in practice would make the interaction negligible. In view of these observations, a heavy field is introduced. This field acts as a mediator for the inflaton decay into light fields. This mechanisms results in a two step process of radiation production, $\phi\rightarrow \chi\rightarrow \tilde{y}\tilde{y}$, where $\phi$ represents the inflaton, $\chi$ the intermediary field and $\tilde{y}$ the light sector, composed of fermions $\Phi_y$ and scalars $y$. Also, in order to keep the flatness of the potential, one can resort to work within the framework of supersymmetry, since with such scenario, quantum fluctuations from fermions and bosons cancel out, which is a welcome feature to avoid spoiling the slow-roll conditions that are necessary for the flatness of the potential.

Once radiation is present during the whole inflation epoch, it is important to compute thermal corrections coming from the light particles to verify whether in a finite temperature environment the slow-roll conditions are still maintained.  It has been shown that in the context of supersymmetry \cite{HM, Mar1} and natural inflation \cite{Mishra-Visinelli}, the quantum and radiative corrections do not spoil the slow-roll conditions required for inflation.

We start by briefly describing the supersymmetric model used in Ref. \cite{HM} and the results when considering only  finite temperature corrections.

Within the above mentioned conditions, the superpotential is
\be
W=g\Phi \Lambda^2 -g\Phi X^2,
\label{superpotential}
\ee
where the scalar components of the chiral superfields $\Phi$ and $X$ are $\varphi$ and $\chi$, respectively. $\Lambda$ is a constant. The scalar interaction terms are derived from the superpotential in Eq. (\ref{superpotential}) as
\be
\mathcal{L}_S = -|\partial_\Phi W|^2 - |\partial_X W|^2.
\ee
Defining $\phi=\sqrt{2}\ {\mbox{Re}}(\varphi)$, the inflaton potential, up to one loop coprrection is
\be
V(\phi)=\frac{1}{2}g^2 M_s^2 \left[\phi^2 \ln\left(\frac{\phi^2}{\phi_0}\right)+\phi^2_0-\phi^2\right],
\ee
where $\phi_0$ is the vev of the inflaton field.

To complete the two stage decay process, a light sector $Y$ is introduced and its interaction with the heavy field $X$ is given by:
\be
W_{\mbox{\tiny{light}}}= -hXY^2.
\ee
Furthermore, a Yukawa sector is added to represent the interaction between the scalar and the fermionic sector:
\be
\mathcal{L}_{\mbox{\tiny{Yukawa}}}=-\frac{1}{2}\frac{\partial^2 W}{\partial \phi_n\partial \phi_m}\bar \psi_n P_L \psi_m
-\frac{1}{2}\frac{\partial^2 W^*}{\partial \phi^*_n\partial \phi^*_m}\bar\psi_n P_R \psi_m,
\ee
where $\phi_m$ is a superfield and
\be
   P_{L,R}=(1\mp\gamma_5)/2.
\ee

The quantum corrections to the inflaton potential are shown to be small due to fermion-boson cancellations. The contribution from thermal corrections to the inflaton mass coming from heavy sector loops is Boltzmann suppressed. These $X$ fields are too heavy to be produced on shell and only appear as virtual $\chi$ (bosons) and $\Psi_\chi$ (fermions) pairs, that decay into the light fields. The heavy fields also decay to inflaton particles through $\chi \rightarrow y y \phi$ but this is a sub-leading process compared to $\chi \rightarrow \tilde y \tilde y$, \cite{being-warm}.  Furthermore, it is assumed that there is a soft SUSY breaking in the heavy sector and that light radiation thermalizes.	

We are interested in the full set of interactions that involve the inflaton and the $\chi$ field, that can be read off from the scalar (${\cal L}_s$) and fermion (${\cal L}_f$) sectors of the Lagrangian, given by
\begin{eqnarray} \nonumber
{\cal L}_s & = &g^2 |\Lambda^2-|\chi|^2|^2 +4g^2 |\varphi|^2|\chi|^2\\
&+&4h^2|y|^2|\chi|^2+h^2|y|^4
+2gh(y^2\varphi^\dagger\chi^\dagger+y^{\dagger2}\varphi\chi)\\   \nonumber
{\cal L}_f &= &g(\varphi \overline \psi_{\chi} P_L\psi_{\chi}
+\varphi^\dagger \overline \psi_{\chi}P_R \psi_{\chi})\\ \nonumber
&+&h(\chi \overline\psi_{y}
P_L\psi_{y} + \chi^\dagger \overline\psi_{y} P_R \psi_{y})\\ \nonumber
&+&2g(\chi\overline\psi_{\chi}P_L \psi_{\varphi} +
\chi^\dagger\overline\psi_{\chi} P_R \psi_{\varphi}) \\
&+&2h(y\overline\psi_{y} P_L\psi_{\chi} + y^\dagger\overline\psi_{y}
P_R\psi_{\chi}),
\label{lagrangian}
\end{eqnarray}
where $y$ is the scalar field component of the chiral  superfield $Y$,  $\Psi_i$ denotes the fermion fields coming from the different sectors, $g$ and $h$ are coupling constants and $\Lambda$ is a mass scale. Normalizing the density perturbation amplitude to the cosmic microwave background leads to coupling constants $g$ and $h\sim 0.1$ and a mass scale of up to $\Lambda\sim 10^{11} GeV$.

%%%%%%%%%%%%%%%%%%%%%%%%%%%%%%%%%%%%
\begin{figure}
  \includegraphics[width=5cm]{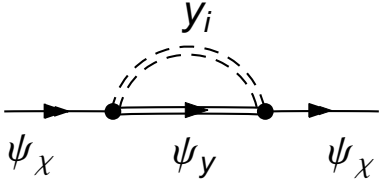}
  \caption{Feynman diagram for the heavy fermion sector
    self-energy. Double line indicates particles can couple to
    an external magnetic field.}
  \label{Fig:fermionselfenergy}
\end{figure}
%%%%%%%%%%%%%%%%%%%%%%%%%%%%%%%%%%%%

Since the only particles that thermalize are the light ones, namely $\tilde{y}$, the thermal effects come from the $\chi$ and $\Psi_{\chi}$ self-energies, containing $\tilde{y}$ propagators. These self-energies can in turn be computed by means of a Hard Thermal Loop (HTL) approximation (see e.g. \cite{Bellac96}). Loops containing heavy particles are suppressed. The external particles can either be $\chi$ or $\Psi_{\chi}$ fields and their masses satisfy $T \ll m_{\chi}, m_{\Psi_{\chi}}$. Therefore, the only Feynman diagrams that need be computed to include both the thermal and the magnetic corrections to the boson and fermion masses are the ones depicted in figs.~\ref{Fig:fermionselfenergy} and~\ref{Fig:bosonselfenergy}.

\subsection{Thermal contribution}

Working in the imaginary-time formalism of thermal field theory and adopting the notation where four (three)-momenta are written in upper (lower) case letters, the fermion self-energy, corresponding to fig.~\ref{Fig:fermionselfenergy} is obtained \cite{HMB, Bellac96} from
\begin{eqnarray}
\label{3momenta}
\Sigma(P) = -4 h^2\,T\,\sum_n\int \frac{d^3k}{(2 \pi)^3}
(K\!\!\!\!/ - P\!\!\!\!/) \Delta(K) \widetilde{\Delta}(P-K),\nonumber\\
\label{ferself}
\end{eqnarray}
where $\Delta$ and $\tilde{\Delta}$ denote boson and fermion propagators, respectively. $\Delta(K)\approx K^{-2}$. $k^0=2n\pi T$ is a boson whereas $k^0=(2n+1)\pi T$ corresponds to a fermion Matsubara frequency, respectively. In the infrared limit ($p_0=0$, $p \rightarrow 0$), Eq.~(\ref{ferself}) leads to
\begin{equation}
m_f^2 \equiv \Sigma \approx \frac{h^2 T^2}{2}.
\end{equation}

%%%%%%%%%%%%%%%%%%%%%%%%%%%%%%%%%%%%
\begin{figure}
\begin{center}
\begin{tabular}{ccc}
  \includegraphics[width=5cm]{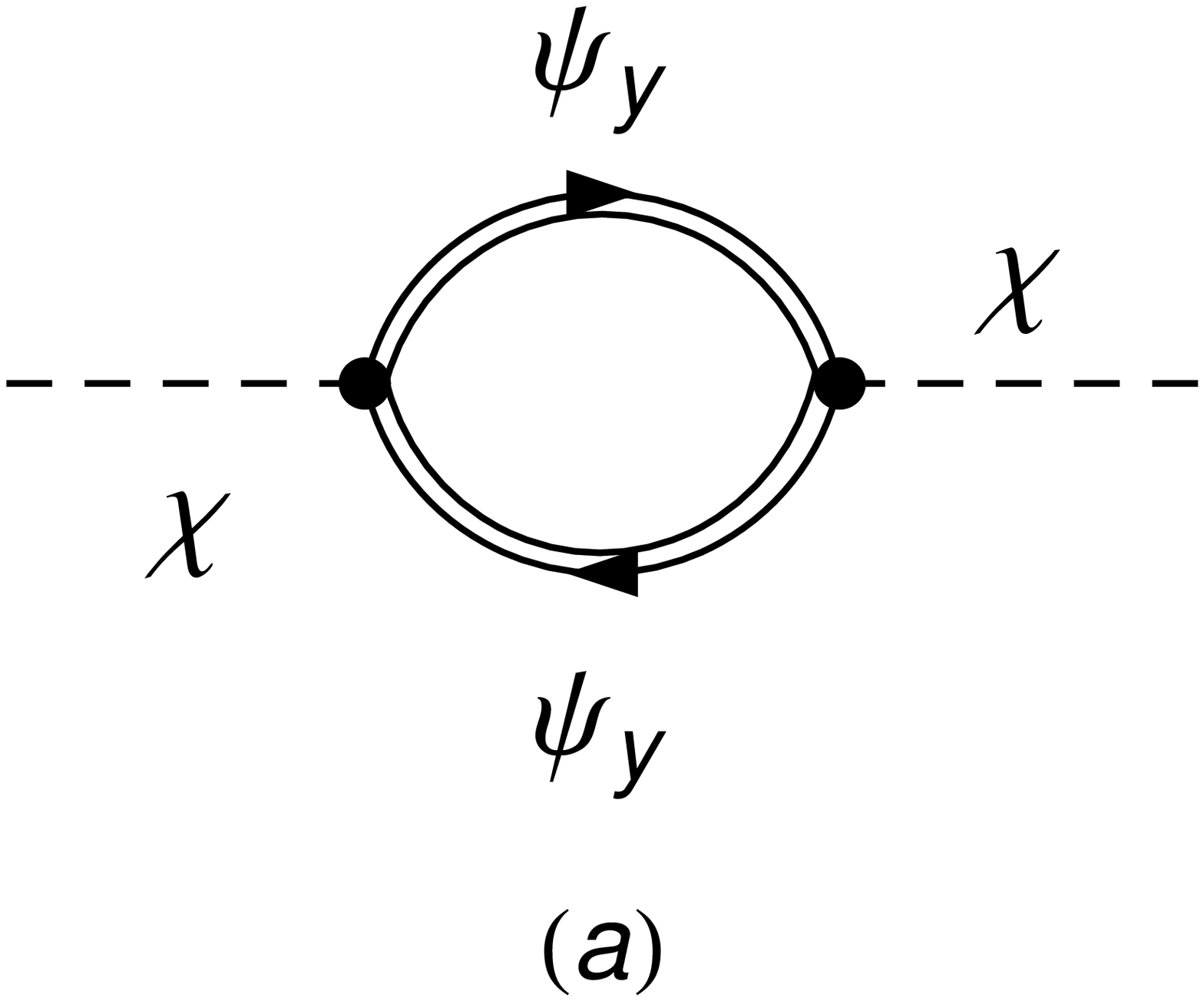}\\
  \\
    \includegraphics[width=5cm]{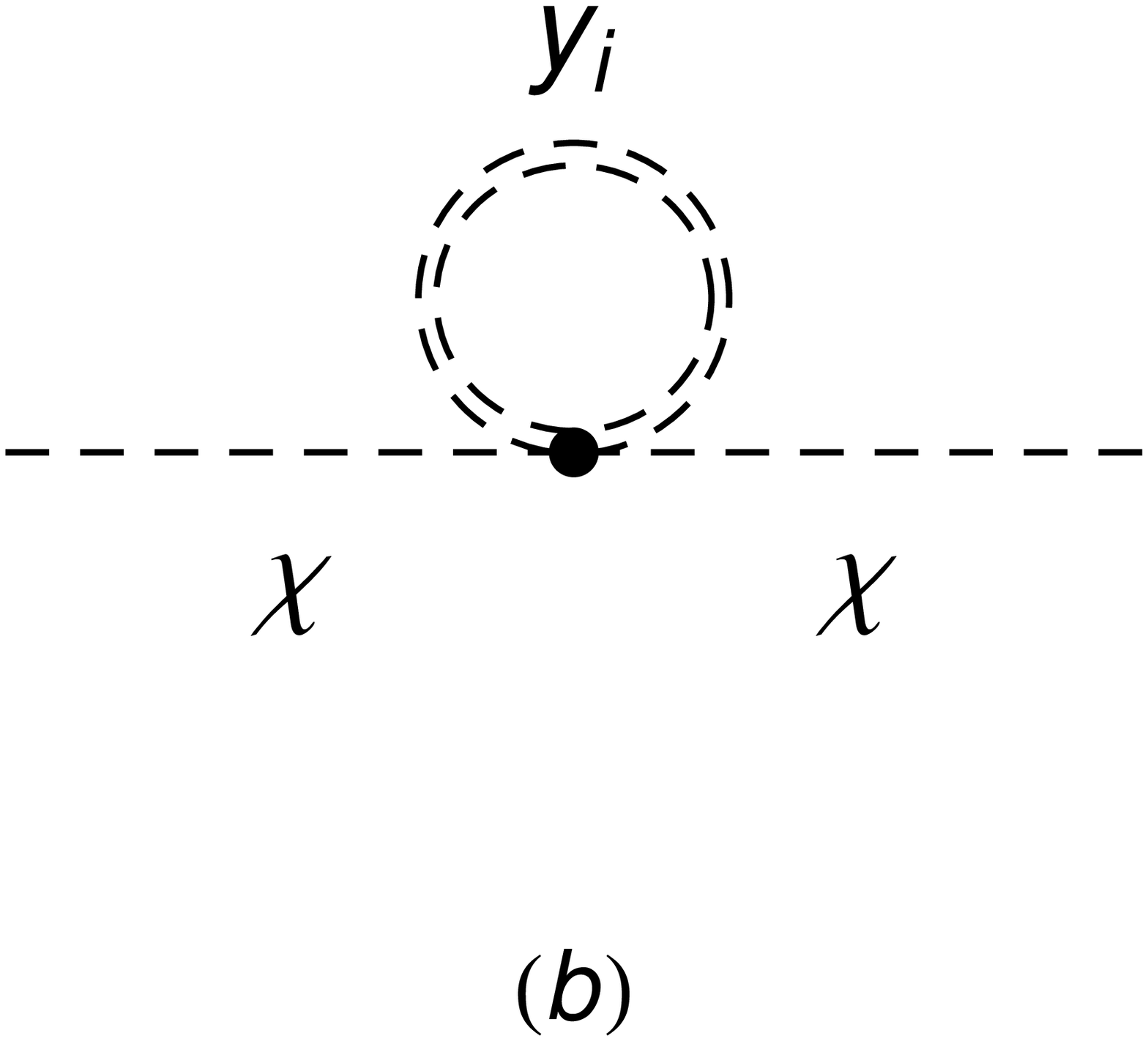}\\
    \\
      \includegraphics[width=5cm]{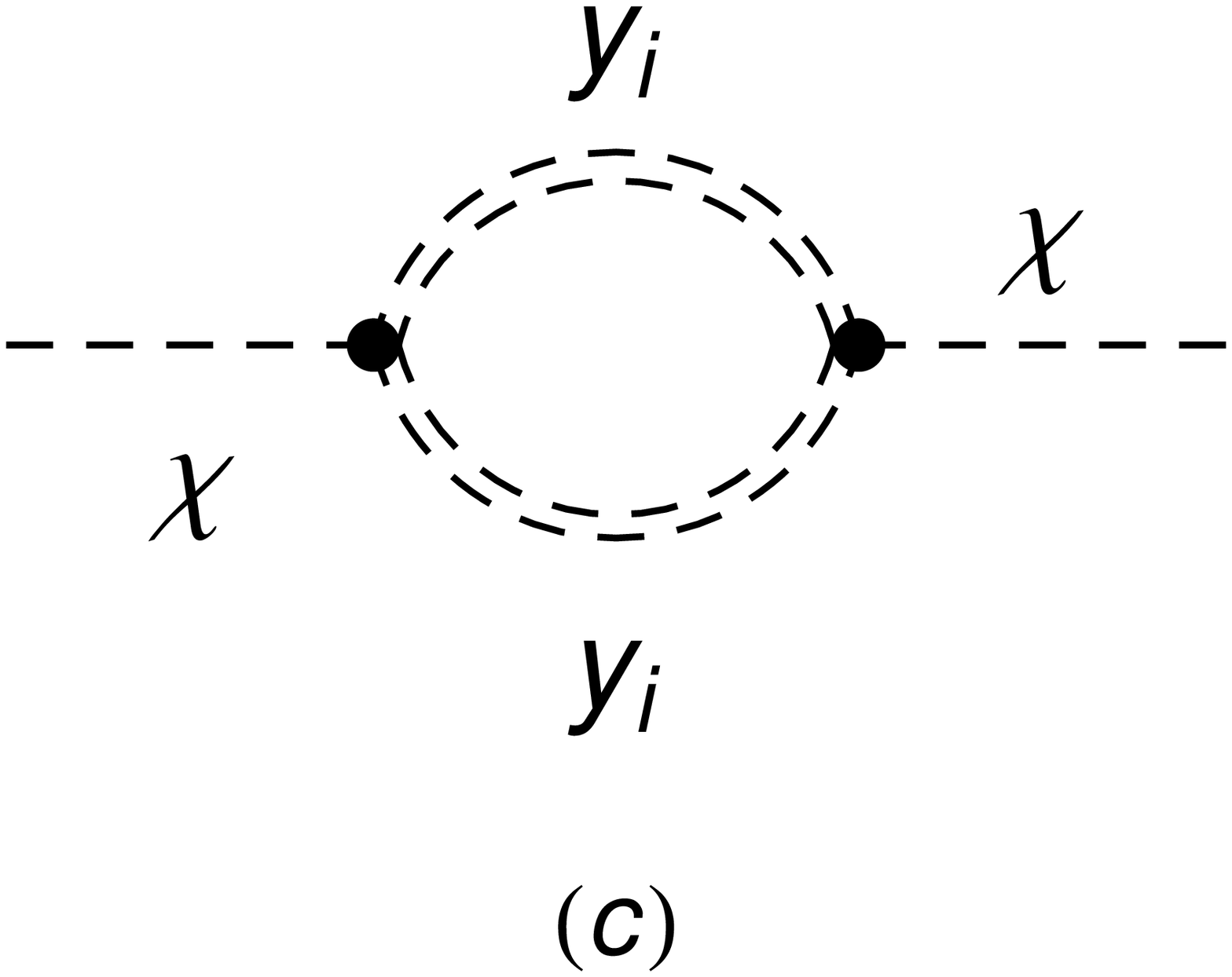}
  \end{tabular}
    \end{center}
    \caption{Feynman diagram for the heavy boson sector self-energy. Double line indicates particles can couple to
  an external magnetic field. }
     \label{Fig:bosonselfenergy}
\end{figure}
%%%%%%%%%%%%%%%%%%%%%%%%%%%%%%%%%%%%

The contributions from the three diagrams in fig.~\ref{Fig:bosonselfenergy} are given  explicitly by
\begin{equation}
\Pi(P)_a = h^2\,T\,\sum_n\int \frac{d^3k}{(2 \pi)^3}
{\mbox{Tr}}\left[K\!\!\!\!/(K\!\!\!\!/ - P\!\!\! \!/)\right]
\widetilde{\Delta}(K) \widetilde{\Delta}(K-P),
\end{equation}
which in the HTL limit becomes

\bea
\label{HTLeq}
\Pi(P)_a &=& -4 h^2 \,T\,\sum_n\int \frac{d^3k}{(2 \pi)^3}K^2
\widetilde{\Delta}(K) \widetilde{\Delta}(K-P)\nonumber\\ 
&\approx& {1 \over 6}h^2 T^2.
\eea

Similarly,
\bea
\label{finitetempself}
\Pi(P)_b &=& 4 h^2\,T\,\sum_n\int \frac{d^3k}{(2 \pi)^3}
{\Delta}(K) \nonumber\\
&\approx& \frac13h^2T^2,
\eea

and

\bea
\Pi(P)_c &=& 4g^2 h^2\phi^2\,T\,\sum_n\int \frac{d^3k}{(2
  \pi)^3}{\Delta}(K) {\Delta}(K-P) \nonumber \\
% &\approx& {1\over 2\pi^2}g^2 h^2\phi^2\ln{T^2\over p^2}.
\eea

The contribution from this last self-energy is subdominant for the case where $T$ is the largest of the scales and hereafter we discard it.
On the other hand, the contribution from diagrams $(a)$ and $(b)$ define the thermal correction to the boson mass

\begin{equation}
m^2_b \equiv \Pi_a + \Pi_b \approx \frac{h^2 T^2}{2}.
\end{equation}

In addition to thermal corrections, primordial magnetic fields can also modify the thermodynamic potential. In the next section we compute thermo-magnetic corrections, adopting the same approach as in Ref. \cite{HM} where only thermal contributions were considered. We use the Schwinger proper-time method to account for the influence of a magnetic background.

\section{Thermo-magnetic contribution}

\subsection{Self-energies}

In order to preserve invariance under $U(1)$ transformations, the only charged fields in the model can be $y$ and $\psi_y$. The magnetic corrections are obtained from the loop corrections to the propagators of the heavy fields $\chi$ and $\psi_\chi$, in the same way that the thermal corrections were obtained. We work with a constant magnetic field of strength $B$ along the $z$ axis and with the assumption that the hierarchy of scales $eB <m_y^2<<T^2$ is obeyed, where $m_y$ is the mass of the fields inside the loop.

To include the effect of an external magnetic field, we use Schwinger's proper-time method \cite{Schwinger51}, where the momentum dependent propagators for charged scalars and fermions coupled to the external field take the form
\begin{eqnarray}
\label{scalpropmom}
D_{B} (k) &=& \int_0^\infty \frac{ds}{\cos{eBs}} \nonumber \\
&& \times \exp\left\{ i s (k_{||}^2-k_{\bot}^2 \frac{\tan{eBs}}{eBs}-m^2_b +i
\epsilon)\right\},\nonumber\\
\end{eqnarray}
and
\begin{eqnarray}
S_{B} (k)&=&\int_0^\infty \frac{ds}{\cos{eBs}} \nonumber \\
& & \times \exp\left\{ i s (k_{||}^2-k_{\bot}^2 \frac{\tan{eBs}}{ eBs}-m^2_{f}
+i \epsilon)\right\} \nonumber \\
& & \times \left[ (m_f+{\not \! k}_{||})e^{i eB s \sigma_3} -\frac{{\not \!
k_\bot}}{\cos{eB s}}\right],\nonumber\\
\label{ferpropmom}
\end{eqnarray}
respectively. We have adopted the notation $k_{||}^2=k_0^2-k_3^2$, $k_\perp^2=k_1^2+k_2^2$, and $\sigma^3=i\gamma^1\gamma^2=-\gamma^5{\not \! u}{\not \! b}$, where $\not \! u$ and $\not \! b$ are four-vectors describing the plasma rest frame and the direction of the magnetic field, respectively.

It has been shown that, by deforming the contour of integration,
Eqs.~(\ref{scalpropmom}) and~(\ref{ferpropmom}) can be written
as \cite{Ayala05,Chyi}
\begin{eqnarray}
D_{B}(k)=2i\sum_{l=0}^{\infty}\frac{(-1)^lL_l(\frac{2k_\perp^2}{e B})
{\mathrm e}^{-\frac{k^2_\perp}{e B}}}{k^2_{||}-(2l+1)e B-m^2_b+i\epsilon},
\label{scalpropsum}
\end{eqnarray}
\begin{eqnarray}
S_{B}(k)= i \sum^\infty_{l=0} \frac{d_l(\frac{k_\perp^2}{e B})D +
d^{\prime}_l(\frac{k_\perp^2}{e B}) \bar D}{k^2_{||}-2 l e B-m_f^2 +
i\epsilon} + \frac{{\not \! k_{\bot}}}{k^2_\perp}, \label{ferpropsum}
\end{eqnarray}
where $d_l(\alpha)\equiv (-1)^n e^{-\alpha} L^{-1}_l(2\alpha)$,
$d^{\prime}_n=\partial d_n/\partial \alpha$,
\begin{eqnarray}
D &=& (m_f+{\not \! k_{||}})+ {\not \! k_{\perp}} \frac{m_f^2-k^2_{||}}{ {
k^2_{\perp}}}, \nonumber \\
\bar D &=& \gamma_5 {\not \! u}{\not \! b}(m_f + {\not \! k_{||}}),
\label{DDe}
\end{eqnarray}
and $L_l$, $L_l^m$ are Laguerre and associated Laguerre polynomials, respectively. Performing a weak field expansion in Eqs.~(\ref{scalpropsum}) and (\ref{ferpropsum})~\cite{Ayala05,Chyi} it is possible to carry out the sum over Landau levels to write the scalar and fermion propagators as power series in $eB$, which up to order $(eB)^2$ read as
\begin{eqnarray}
\label{scalpropweak}
D_{B}(k)&=&\frac{i}{k^2-m^2_b}-(eB)^2\left(\frac{i}{(k^2-m^2_b)^3}
+ \frac{2ik_\perp^2}{(k^2-m^2_b)^4}\right)\nonumber\\
            &\equiv&D_0(k)+(eB)^2 D_2(k)
\end{eqnarray}

\begin{eqnarray} \nonumber
{S_{B}(k)} &=& i\frac{{\not \! k}+m_f}{{k}^2-m_f^2}+
i\frac{\gamma_5 {\not \! u}{\not \! b}(k_{||}+m_f)(eB)}{(k^2-m_f^2)^2}\nonumber\\
&-&i\frac{2(eB)^2 k_\perp^2}{(k^2-m_f^2)^4} (m_f+{\not \! k_{||}}+{\not \!
k_\perp}\frac{m_f^2-k_{||}^2}{k_\perp^2})\nonumber\\
&\equiv& S_0(k)+(eB)\ S_1(k)+(eB)^2S_2(k).
\label{ferpropweak}
\end{eqnarray}

Using these propagators to compute the self-energies, Eqs.~\eqref{3momenta},~\eqref{HTLeq} and~\eqref{finitetempself}, we obtain the leading order corrections, from thermal and magnetic effects, to the heavy sector boson and fermion masses which, as we show in Appendices A and B, can be written as
\begin{eqnarray} \nonumber
m^2_b (T,B) &\approx & %\frac{h^2 T^2}{2}
m_b^2\Bigg( 1 - \frac{2 m_y}{\pi T}
- \frac{m^2_y}{2 \pi^2 T^2} \Big[\ln \left( \frac{m^2_y}{(4 \pi T)^2} \right)\\
&+&2 \gamma_{_E}- 1 \Big] - {1 \over 12 \pi} \frac{(eB)^2}{m^3_y T} \Bigg)
\label{eq:bosonselfenergy}
\end{eqnarray}
\begin{equation}
m^2_f (T,B;r) \approx %\frac{h^2 T^2}{2}
m_f^2
\left( 1 - {1 \over 3} \frac{r (eB)}{\pi m_y T} + {11 \over 12 \pi} \frac{(eB)^2}{m^3_y T} \right),
\label{eq:fermionselfenergy}
\end{equation}
where $r= \pm 1$ represents the two possible spin orientations of the fermion with respect to the magnetic field.

These masses are then used to correct the fermion and boson propagators
\begin{equation}
S^{-1} \sim P^2 + m^2_{\Psi_\chi},
\label{propagador1}
\end{equation}
\begin{equation}
G^{-1} = P^2 + m^2_{\chi},
\label{propagador2}
\end{equation}
which in turn modify the effective inflaton potential
\begin{eqnarray}
V_\chi &=&\int \frac{d^4P}{(2\pi)^4} \ln\det(G^{-1})\nonumber\\
&-&\int \frac{d^4P}{(2\pi)^4} \ln \det(S^{-1}S^{*-1})^{-1/2}.
\label{potencial1}
\end{eqnarray}

%Furthermore, the boson and fermion masses differ not only because their magnetic field dependence is different, but also due to the introduction a soft SUSY breaking term which contributes with $M_s^2$ to the square of the boson mass~\cite{HM}
%\begin{eqnarray}
%m^2_\chi(T,B)&=&2g^2\phi^2+ m_b^2(T,B) + M_s^2, \nonumber \\
%m^2_{\Psi_\chi}(T,B;r)&=&2g^2\phi^2 + m_f^2(T,B;r).
%\label{masas}
%\end{eqnarray}

%The complete one-loop inflaton potential, including themal and magnetic effects, can be written as
%\begin{equation}
%V(\phi,T, B) = -\frac{\pi^2}{90} g_{*}T^4 + V_\chi(\phi, T, B),
%\end{equation}
%where, as we show in Appendix II, $V_\chi$ is given in the weak field limit by

%\bea
% &&\hspace{-1cm}V_\chi(\phi, T, B)=\frac{M_s^2\left(m_{\Psi _\chi}^2 + m_b^2(T,B)\right)}{16\pi^2}
%\nonumber \\
%&&\hspace{-.5cm}\times\left(1-\sum_{r=\pm1}\frac{m_f^2(T,B;r)-m_b^2(T,B)}{2M_s^2}\right)
%\nonumber \\
%&&\hspace{-0.5cm}\times\left(\ln\left(\frac{m_{\Psi_\chi}^2+m_b^2(T,B)}{{m_0}_{\Psi_\chi}^2}\right)-1\right)+\frac{M_s^2{m_0}_{\Psi_\chi}^2 }{16\pi^2},
%\label{Vchi}
%\eea
%with ${m_0}_{\Psi_\chi}\equiv \sqrt{2}g\phi_0$.

\subsection{Effective potential}

The boson and fermion masses differ not only because their magnetic field dependence is different, as given by Eqs.~(\ref{eq:bosonselfenergy}) and~(\ref{eq:fermionselfenergy}), but also due to the introduction of a soft SUSY breaking term which contributes with $M_s^2$ to the square of the boson mass~\cite{HM}
\begin{eqnarray}
m^2_\chi(T,B)&=&2g^2\phi^2+ m_b^2(T,B) + M_s^2, \nonumber \\
m^2_{\Psi_\chi}(T,B;r)&=&2g^2\phi^2 + m_f^2(T,B;r).
\label{eq:susybreaking}
\end{eqnarray}

The complete one-loop inflaton potential, including thermal and magnetic effects, can be written as
\begin{equation}
V(\phi,T, B) = -\frac{\pi^2}{90} g_{*}T^4 + V_\chi(\phi, T, B),
\label{potencial2}
\end{equation}
where $V_\chi(\phi, T, B)$ is made of a boson and a fermion pieces, that can be evaluated by means of Eqs.~(\ref{potencial1}). To deal with the infinities that these integrals contain, we introduce an ultraviolet cutoff $\Lambda_{uv}$. Due to supersymmetry the boson and fermion terms proportional to the higher powers of $\Lambda_{uv}$ cancel out
and, as we show in Appendix B, the total potential $V_\chi$ is given in the weak field limit by

\bea
 &&\hspace{-1cm}V_\chi(\phi, T, B)=\frac{M_s^2\left(m_{\Psi _\chi}^2 + m_b^2(T,B)\right)}{16\pi^2}
\nonumber \\
&&\hspace{-.5cm}\times\left(1-\sum_{r=\pm1}\frac{m_f^2(T,B;r)-m_b^2(T,B)}{2M_s^2}\right)
\nonumber \\
&&\hspace{-0.5cm}\times\left(\ln\left(\frac{m_{\Psi_\chi}^2+m_b^2(T,B)}{{m_0}_{\Psi_\chi}^2}\right)-1\right)+\frac{M_s^2{m_0}_{\Psi_\chi}^2 }{16\pi^2},
\label{eq:Vchi}
\eea
where we have used Eq.~(\ref{eq:susybreaking}) and have defined ${m_0}_{\Psi_\chi}\equiv \sqrt{2}g\phi_0$.

From Eq.~(\ref{eq:Vchi}) one can notice that in the absence of supersymmetry breaking and considering only the thermal contribution, the potential vanishes. This happens because in such a case the fermion and boson masses, $m_{\psi_\chi}$ and $m_\chi$, are the same and in the HTL approximation the self-energies from where these masses are obtained, become equal. On the other hand, the magnetic contribution tends to break supersymmetry.

%%%%%%%%%%%%%%%%%%%%%%
\begin{figure}[t!]
\vspace{0.42cm} {\centering
\resizebox*{0.42\textwidth}
{0.22\textheight}{\includegraphics{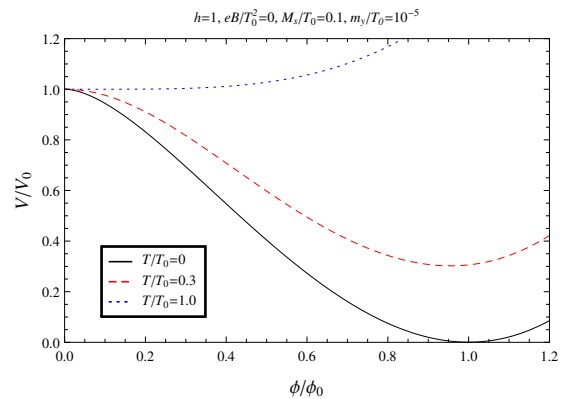}}
\par}
\caption{Color on-line. Effective potential as a function of $\phi/\phi_0$ at zero magnetic field for three different temperatures scaled to $T_0$, the critical temperature for the phase transition to take place in the absence of the magnetic field. Note that as the temperature increases symmetry is restored.}
  \label{fig3}
\end{figure}
%%%%%%%%%%%%%%%%%%%%%%
%%%%%%%%%%%%%%%%%%%%%%
\begin{figure}[b!]
\vspace{0.42cm} {\centering
\resizebox*{0.42\textwidth}
{0.22\textheight}{\includegraphics{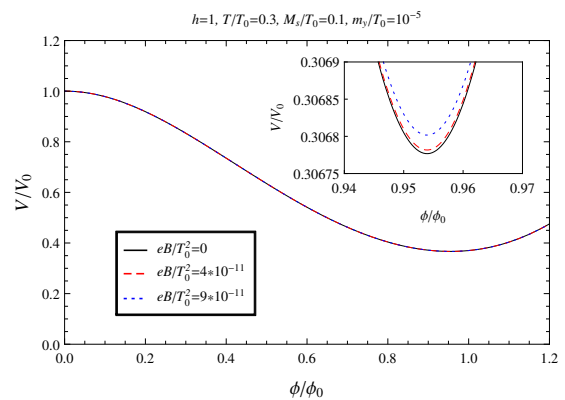}}
\par}
\caption{Color on-line. Effective potential as a function of $\phi/\phi_0$ for different values of the magnetic field strength and a fixed temperature. Note that as the magnetic field increases, the effective potential becomes shallower and thus the phase transition is delayed.}
   \label{fig4}
\end{figure}
%%%%%%%%%%%%%%%%%%%%%%

%%%%%%%%%%%%%%%%%%%%%%
\begin{figure}[t!]
\vspace{0.42cm} {\centering
\resizebox*{0.42\textwidth}
{0.22\textheight}{\includegraphics{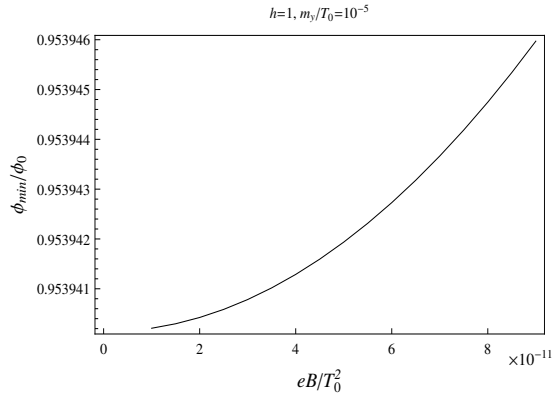}}
\par}
\caption{$\phi_{min}/\phi_0$ as a function of the the magnetic field. The minimum in the broken phase grows with the field strength and the curvature of the effective potential near the origin becomes flatter.}
\label{fig5}
\end{figure}
%%%%%%%%%%%%%%%%%%%%%%
%%%%%%%%%%%%%%%%%%%
\begin{figure}[b!]
\vspace{0.42cm} {\centering
\resizebox*{0.42\textwidth}
{0.22\textheight}{\includegraphics{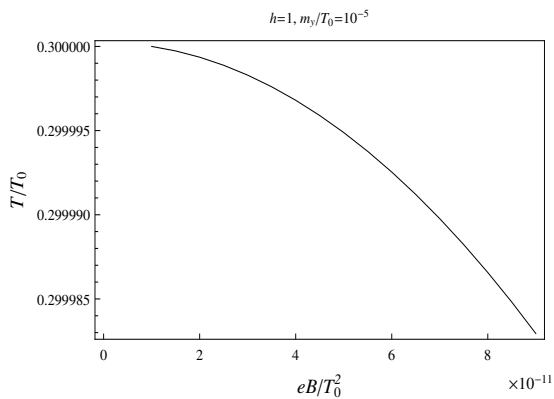}}
\par}
\caption{Temperature where the minima of the effective potential in the broken symmetry phase coincide, as a function of the magnetic field.}
  \label{fig6}
\end{figure}
%%%%%%%%%%%%%%%%%%%

Figure~\ref{fig3} shows the effective potential normalized by
$V_0\equiv V(0,0,0)$ as a function of $\phi/\phi_0$ in the absence of magnetic field for three different temperatures scaled to $T_0$, the critical temperature for the phase transition to take place in the absence of the magnetic field. We note that the effective potential shows the expected thermal properties; as temperature increases symmetry is restored.

Figure~\ref{fig4} shows the effective potential as a function of $\phi/\phi_0$ for different values of the magnetic field strength scaled to $T_0^2$ and a fixed temperature corresponding to the broken symmetry phase. Note that the magnetic field effects are small, since we are considering a very weak field, however, as can better be seen in the figure's inset, with increasing magnetic field the effective potential becomes shallower which means that the inflaton's potential becomes a bit flatter and therefore that the phase transition is delayed since this last requires a lower temperature to take place.

Figure~\ref{fig5} shows the behavior of $\phi_{min}/\phi_0$  as a function of the magnetic field. Note that the minimum in the broken phase grows as a function of the field strength. Here, the temperature is decreasing in order to allow the transition to develop. This behavior also implies that the curvature of the effective potential near the origin becomes flatter. That the magnetic field delays the phase transition is also shown in fig.~\ref{fig6} where we plot the behavior of the temperature where the minima of the effective potential in the broken symmetry phase coincide, as a function of the magnetic field. This temperature is a decreasing function of the field strength.

%%%%%%%%%%%%%%%%%%%
\begin{figure}[t!]
\vspace{0.42cm} {\centering
\resizebox*{0.42\textwidth}
{0.22\textheight}{\includegraphics{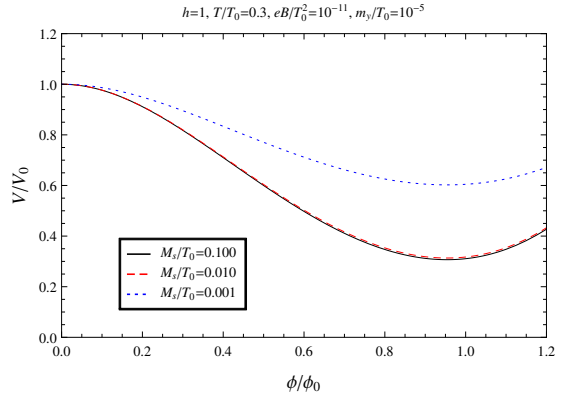}}
\par}
\caption{Color on-line. Effective potential as a function of the soft SUSY-breaking parameter $M_s$. As $M_s$ decreases,  symmetry is restored.}
  \label{fig7}
\end{figure}
%%%%%%%%%%%%%%%%%%%

Figure~\ref{fig7} shows the behavior of the effective potential as we vary the soft SUSY-breaking parameter $M_s$. Note that, as expected, when $M_s$ decreases, symmetry is restored. The effect of the mass of the light particles on the effective potential is shown in fig.~\ref{fig8}. Symmetry tends to also be restored as the mass of the light particles increases.

\section{Summary and Conclusions}

In this work we have studied the effects that a possible primordial magnetic field can have on the inflation's potential, taking as the underlying model a warm inflation scenario. The model is based on global supersymmetry and a coupling between the inflaton and heavy intermediate superfields which are in turn coupled to light particles. The inflaton decay is therefore a two step process and the effects of the magnetic field are felt by the light charged particles which produce a modification to the masses of the heavy intermediate fields that receive thermo-magnetic corrections. We have shown that the presence of the magnetic field delays the phase transition and makes the effective potential flatter than when considering purely thermal corrections, rendering the transition smoother.

The working assumption is a simple scenario whereby the magnetic field scales adiabatically with the universe expansion from the end of the inflationary epoch. However, to be on the safe side and consider that the field strength is probably the smallest of the energy scales at that epoch, we have adopted a conservative scenario where the intensity of the field and the temperature are related by $eB = a T^2$, where $a \ll 1$. This approach should allow some room to include scenarios where the observed field in the current epoch has been amplified for instance by a field helicity, and thus corresponds to a weaker field during inflation than the one obtained by a simple adiabatic expansion. Other kinds of evolution discussed in the literature are possible such as an anisotropic expansion driven by the magnetic field itself, the interpretation of the progenitor magnetic field as a proper average of small flux elements which depends on the properties of the random fields when cosmic magnetic fields are tangled on scales smaller than the observed ones, the enhancing of the magnetic field induced by the inverse cascade phenomenon driven in turn by a magnetic helicity and the phenomenon of back reaction to electromagnetic fields during inflation (see for instance Refs.~\cite{Berera1999, Yamazaki12, Brandenburg96, Thorne1967, Fujita2014}). 

The generation of the widespread magnetic fields in the universe is an open problem. Some works report on the possibility of effectively generating observationally interesting large-scale magnetic fields during inflation (see e.g. Refs.~\cite{Berera1999,Subramanian2010,Martin2008}). In particular, in Ref.~\cite{Berera1999}, the authors work in the context of warm inflation, and consider the possible role of helicity for magnifying magnetic seeds emerging from a Savvidy vacuum scenario. Nevertheless, we stress that the magnetogenesis mechanism is not the central problem in our work. Our main drive is to consider the observed field magnitudes at present to get a simple upper bound for the strength of the primordial fields during inflation and thus to have a definite hierarchy of scales. The application of these ideas to a particular model is for the time being outside the scope of this work where the main focus is to present a general scenario and to show the modifications that a magnetic field can induce on the inflationary potential within the context of warm inflation. 

%%%%%%%%%%%%%%%%%%%
\begin{figure}[t!]
\vspace{0.42cm} {\centering
\resizebox*{0.42\textwidth}
{0.22\textheight}{\includegraphics{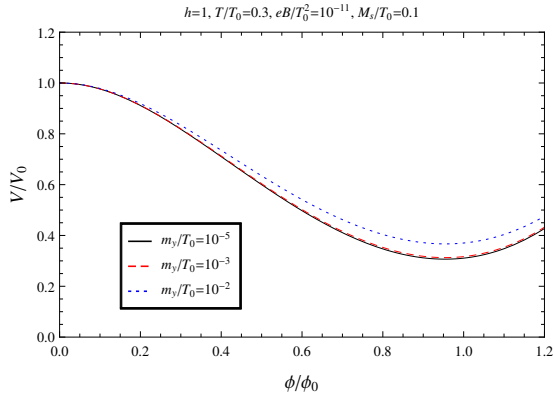}}
\par}
\caption{Color on-line. Effective potential as a function of $\phi/\phi_0$ for different values of the light particle's mass. As the mass increases, symmetry tends to be restored.}
  \label{fig8}
\end{figure}
%%%%%%%%%%%%%%%%%%%

Although the flatness of the potential is not spoiled by the presence of the magnetic field, more detailed and systematic studies of the effects on the slow roll conditions as well as on dissipation and estimates within particular models, need to be carried out. This work is on its way and it will be reported elsewhere.

\section*{Acknowledgments}

A.J.M is in debt to R. O. Ramos and G.P. to M. Bastero-Gil and J.G. Rosas for helpful discussions. Support for this work has been received in part from DGAPA-UNAM under grant numbers PAPIIT-IN103811, PAPIIT-IN117111, PAPIIT-IN117914 and CONACyT-M\'exico under grant number 128534.

\appendix

\section{Femion self-energy}

Here we show the details of the calculation for the fermion
self-energy in Fig. 1, at finite temperature in the presence of an external
magnetic field.

The fermion self-energy in vacuum is given by
\bea
\Sigma=-2h^2\int\frac{d^4k}{(2\pi)^4}
       \left(S_B(p-k)-\gamma_5S_B(p-k)\gamma_5\right)D_B(k),\nonumber\\
\label{selfap1}
\eea
where $D_B(k)$ and $S_B(p-k)$ are the boson and fermion propagators,
respectively. In the weak field approximation, $eB < m^2$, these are given by Eqs.~(\ref{scalpropweak}) and~(\ref{ferpropweak}). Therefore the self-energy can be written as
\bea
   \Sigma^{B}(p)=\Sigma^0(p)+eB\, \Sigma^1(p)+(eB)^2\Sigma^2(p),
\label{selfall}
\eea
where
\bea
   \Sigma^0(p)
     &=&-2h^2\int\frac{d^4k}{(2\pi)^2}\left[S_0-\gamma_5S_0\gamma_5\right]D_0
\label{self00}\\
   \Sigma^1(p)&=&
        -2h^2\int\frac{d^4k}{(2\pi)^2}\left[S_1-\gamma_5S_1\gamma_5\right]D_0
\label{self01}\\
   \Sigma^2(p)
%\nonumber \\
  &=&-2h^2\int\frac{d^4k}{(2\pi)^2}\left[(S_0-\gamma_5S_0\gamma_5)D_2\right.
\nonumber \\
  &&\hspace{2cm}+\left.(S_2-\gamma_5S_2\gamma_5)D_0\right],
%\nonumber \\
\label{self02}
\eea
where we omit to write explicitly the argument of each propagator. Note that
according to \Eq{selfap1}, all fermion propagators have argument
$(p-k)$ and all boson propagators have argument $(k)$.

Using the imaginary-time formalism, the finite temperature
effects in \Eq{selfall} are introduced by the
replacements $k_0 \rightarrow -i\omega_n$ with $\omega_n=2 n\pi T$ for
bosons and $\omega_n=(2n+1)\pi T$ for fermions, together with
\bea
\int\frac{d^4k}{(2\pi)^4}\rightarrow \sumint\frac{d^4K}{(2\pi)^4} \equiv T \sum_{n=-\infty}^\infty \int \frac{d^3k}{(2\pi)^3}.
\eea
Hereafter we calculate each term in \Eq{selfall} in the HTL approximation. Let us start by calculating the first term in \Eq{selfall} which is written as
\bea
\hspace{-0.7cm} \Sigma^0_T=-4h^2 \sumint\frac{d^4K}{(2\pi)^4}
(\slsh{K}-\slsh{P})\widetilde{\Delta}(P-K)\Delta(K),
%\frac{}{(p-k)^2+m_{\psi_y}^2}\frac{1}{k^2+m_{y}^2}
\label{self00T}
\eea
with $\widetilde{\Delta}(K)\equiv (K^2+m_{\psi_y}^2)^{-1}$ and  ${\Delta}(K)\equiv (K^2+m_{y}^2)^{-1}.$
After performing the sum over the Matsubara frequencies and the
integral over the solid angle in \Eq{self00T},  we get
\bea
\hspace{-0.7cm}\Sigma^0_T=-4h^2
\sum_{r=\pm1}\left(i\gamma_4(I^r_1+I^r_2)-\gamma_i\hat{p}_i (I^r_3-I^r_4)\right),
%\nonumber \\
\label{ferautoT}
\eea
with
\bea
I^r_1&=&\int
\frac{k^2dk}{(2\pi)^2}\frac{n_b(E_1)}{2E_1} r(E_1-rp_0)
\Theta(rE_1,k,1)
\\
I^r_2&=&\int
\frac{k^2dk}{(2\pi)^2}\frac{n_f(E_2)}{2} r
\Theta(rE_2,k,-1)
\\
I^r_3&=&\int
\frac{k^2dk}{(2\pi)^2}\frac{n_b(E_1)}{2E_1}\left( \frac{1}{2p} \right)
\nonumber \\
&&\hspace{-.5cm}\times
\left[2-(-2rE_1p_0+p_0^2+p^2)\Theta(rE_1,k,1)\right]
%\nonumber \\
\\
I^r_4&=&\int
\frac{k^2dk}{(2\pi)^2}\frac{n_f(E_2)}{2E_2}\left( \frac{-1}{2p} \right)
\nonumber \\
&&\hspace{-.5cm}\times
\left[2-(-2rE_2p_0+p_0^2+p^2)\Theta(rE_2,k,-1)\right],
\eea
where $E_1^2\equiv k^2+m_y^2$ and $E_2^2\equiv k^2+m_{\psi_y}^2$
denote the light particles' energies and $n_b(E_1)$ and $n_f(E_2)$ are
the Bose-Einstein and Fermi-Dirac distributions, respectively. The
function $\Theta(rE,k,\alpha)$ is defined as
\bea
   &&\Theta(rE_1,k,\alpha)\equiv
\nonumber \\
&&\frac{1}{2 kp}
    \ln\left[
       \frac{-2rE_1p_0+2kp+p_0^2+p^2+\alpha(m_y^2-m_{\psi_y}^2)}{-2rE_1p_0-2kp+p_0^2+p^2+\alpha(m_y^2-m_{\psi_y}^2)}
       \right].
\nonumber \\
\eea
In the HTL approximation $E\approx k$ and
$n_{f,b}(E)\approx n_{f,b}(k)$, so, each of $I^r_i$ reduces to
\bea
I^r_1&=&\frac{1}{2}I^r_2=-\frac{T^2}{48}\frac{r}{p}Q_0\left(\frac{rp_0}{p}\right)
\label{htlfer1}\\
I^r_3&=&-\frac{1}{2}I^r_4=\frac{T^2}{48}\frac{1}{p}\left(1-\frac{rp_0}{p}Q_0\left(\frac{rp_0}{p}\right)\right),
\label{htlfer2}
\eea
with
\bea
   Q_0(x)=\frac{1}{2}\ln\left(\frac{x+1}{x-1}\right).
\eea
Using \Eq{htlfer1} and \Eq{htlfer2} into \Eq{ferautoT}, the fermion
self-energy becomes
\bea
  \Sigma^0_T&=&-
     \frac{m_f^2}{2p}\left\{i\gamma_4Q_0\left(\frac{p_0}{p}\right)+\gamma_i\hat{p}_i\left[1-\frac{p_0}{p}Q_0\left(\frac{p_0}{p}\right)\right]\right\},
\nonumber \\
\label{ferautoThtl}
\eea
where in order to perform the sum over $r$, we used the property
\bea
   rQ_0(r x) = Q(x)
\eea
and $m_f\equiv \frac{h}{\sqrt{2}}T$ is the thermal correction to the mass
of the heavy fermion $\psi_\chi$.

The second term in \Eq{selfall} reads
\bea
  \Sigma_T^1=4h^2\gamma_5\slsh{u}\slsh{b}
%\nonumber \\
%    &&\hspace{-0.5cm}\times
    \sumint \frac{d^4K}{(2\pi)^4}(\slsh{K}_{||}-{\slsh{P}_{||}})
    \widetilde{\Delta}^2(P-K)\Delta(K),
%\frac{}
%         ({(p-k)^2+m_{\psi_y}^2)^2}\frac{1}{k^2+m_y^2}
\nonumber \\
\eea
which can be easily calculated by noticing that
\bea
  &&\hspace{-1cm}\Sigma_T^1=-4h^2\gamma_5\slsh{u}\slsh{b}
\nonumber \\
    &&\hspace{-0.8cm}\times\frac{\partial}{\partial m_{\Psi_y}^2}
    \sumint \frac{d^4K}{(2\pi)^4}(\slsh{K}_{||}-{\slsh{P}_{||}})
    \widetilde{\Delta}(P-K)\Delta(K).
%    \sumint \frac{d^3k}{(2\pi)^3}
%    \frac{\slsh{k}_{||}-{\slsh{p}_{||}}}
%         {(p-k)^2+m_{\psi_y}^2}\frac{1}{k^2+m_y^2}
%\nonumber \\
\label{self01T}
\eea
By comparing  Eqs.~(\ref{self01T}) and~(\ref{self00T}), it is not difficult to
see that \Eq{self01T} can be rewritten as
\bea
   &&\hspace{-1.2cm}
\Sigma_T^1=-4h^2\gamma_5\slsh{u}\slsh{b}
\nonumber \\
    &&\hspace{-1cm}\times
\sum_{r=\pm1}
\frac{\partial}{\partial m_{\Psi_y}^2}
\left(i\gamma_4(I^r_1+I^r_2)-\gamma_3\hat{p}_3 (I^r_3-I^r_4)\right).
\label{self01T2}
\eea
which in the  HTL approximation becomes
\bea
   \Sigma_T^1&=&-\frac{m_f^2}{\pi m_y T}
     \gamma_5\slsh{u}\slsh{b}
     \left(i\gamma_4-\gamma_3\hat{p}_3\frac{p_0^2+p^2}{2p
       p_0}\right)\frac{p^2p_0}{(p^2-p_0^2)^2}.
\nonumber \\
\eea

The last term of \Eq{selfall} at finite temperature reads
\bea
   &&\hspace{-0.6cm}\Sigma_T^2=-4h^2\sumint \frac{d^4K}{(2\pi)^4}
      \widetilde{\Delta}(P-K)\Delta(K)
%\frac{1}{k^2+m_y^2}\frac{1}{(p-k)^2+m_{\psi_y}^2}
\nonumber \\
       &&\times\left[
   -(\slsh{K}-\slsh{P})\Delta^2(K)% \frac{}{(k^2+m_y^2)^2}
   -2(\slsh{K}-\slsh{P})_\perp\widetilde{\Delta}^2(P-K)
    %\frac{}{((p-k)^2+m_{\psi_y}^2)^2}
    \right.
\nonumber \\
    &&+ 2 K_\perp^2(\slsh{K}-\slsh{P})\Delta^3(K)
      %\frac{}{((p-k)^2+m_{\psi_y}^2)^3}
\nonumber \\
     &&\left.+ 2(P-K)_\perp^2(\slsh{K}-\slsh{P})\widetilde{\Delta}^3(P-K),
      %\frac{}{(k^2+m_y^2)^3}
      \right]
%\nonumber \\
\label{self02T0}
\eea
which can be rewritten as
\bea
   &&\hspace{-0cm}\Sigma_T^2=4h^2\sumint \frac{d^4K}{(2\pi)^4}
    \left[
   \frac{1}{2}\frac{\partial^2}{\partial (m_y^2)^2}(\slsh{K}-\slsh{P})
   \right.
\nonumber \\
       &&\hspace{0.5cm}
      +\frac{\partial^2}{\partial
        (m_{\psi_y}^2)^2}(\slsh{K}-\slsh{P})_\perp
%\nonumber \\
%      &&\hspace{0.5cm}
     +\frac{1}{3}\frac{\partial^3}{\partial (m_{y}^2)^3}
       K_\perp^2(\slsh{K}-\slsh{P})
\nonumber \\
       &&\left.\hspace{0.5cm}
      +\frac{1}{3}\frac{\partial^3}{\partial (m_{\psi_y}^2)^3}
       (P-K)_\perp^2(\slsh{K}-\slsh{P})
      \right]
      \Delta(K)
      %\frac{1}{k^2+m_y^2}
      \widetilde{\Delta}(P-K).
      %\frac{1}{(p-k)^2+m_{\psi_y}^2}
\nonumber \\
\label{self02T01}
\eea
To take care of the angular integration we separate \Eq{self02T01} as follows
\bea
   &&\Sigma_T^2=(\Sigma_T^2)_I + (\Sigma_T^2)_J,
\label{sigmass}
\eea
where
\bea
&&\hspace{-1.5cm}(\Sigma_T^2)_I\equiv
4h^2\sumint \frac{d^4K}{(2\pi)^4}
    \left[
   \frac{1}{2}\frac{\partial^2}{\partial (m_y^2)^2}(\slsh{K}-\slsh{P})
   \right.
\nonumber \\
       &&\hspace{-0.5cm}\left.
      +\frac{\partial^2}{\partial (m_{\psi_y}^2)^2}(\slsh{K}-\slsh{P})_\perp
      \right]
     \Delta(K) \widetilde{\Delta}(P-K)
%      \frac{1}{k^2+m_y^2}\frac{1}{(p-k)^2+m_{\psi_y}^2}
%\nonumber \\
\label{self02T02}
\eea
and
\bea
   &&\hspace{-0.5cm}(\Sigma_T^2)_J\equiv4h^2\sumint \frac{d^4K}{(2\pi)^4}
    \left[
      \frac{1}{3}\frac{\partial^3}{\partial (m_{\psi_y}^2)^3}
       (P-K)_\perp^2(\slsh{K}-\slsh{P})\right.
\nonumber \\
       &&\left.\hspace{0.5cm}
     +\frac{1}{3}\frac{\partial^3}{\partial (m_{y}^2)^3}
       K_\perp^2(\slsh{K}-\slsh{P})
      \right]\Delta(K) \widetilde{\Delta}(P-K).
%      \frac{1}{k^2+m_y^2}\frac{1}{(p-k)^2+m_{\psi_y}^2}
%\nonumber \\
\label{self02T03}
\eea
Since \Eq{self02T02} is similar to \Eq{self00T}, it is simple to see
that  \Eq{self02T02} can be rewritten as
\bea
 (\Sigma_T^2)_I&=&
4h^2\left[
   \frac{1}{2}\frac{\partial^2}{\partial (m_y^2)^2}
   (i\gamma_4(I^r_1+I^r_2)-\gamma_i\hat{p}_i(I^r_3-I^r_4))
   \right.
\nonumber \\
       &&\hspace{1cm}\left.
      +\frac{\partial^2}{\partial (m_{\psi_y}^2)^2}
       (\gamma_i^\perp\hat{p}^\perp_i(I^r_3-I^r_4)),
      \right]
\label{self02T04}
\eea
which in the HTL approximation reduces to
\bea
 (\Sigma_T^2)_I&=&
4h^2\left[
   -\frac{1}{2}i\gamma_4Q_0\left(\frac{p_0}{p}\right)
   \right.
\nonumber \\
   &&\left.-\gamma_i\left(\frac{1}{2}\hat{p}_i-\hat{p}^\perp_i\right)
   \left(2-Q_0\left(\frac{p_0}{p}\right)\right)
    %   &&\hspace{1cm}\left.
      \right]\frac{3\pi T}{32 m_y^3 p}.
\nonumber \\
\label{self02T05}
\eea
In a similar fashion, the second term of \Eq{sigmass} can be rewritten as
\bea
    &&\hspace{-1.5cm}(\Sigma_T^2)_J=4h^2
    \left[
      \frac{1}{3}\frac{\partial^3}{\partial (m_{\psi_y}^2)^3}
     +\frac{1}{3}\frac{\partial^3}{\partial (m_{y}^2)^3})
      \right]
\nonumber \\
      &&\times\sum_{r=\pm1}
         \left[i\gamma_4(J_1^r+J_2^r)-\gamma_i\hat{p}_i(J^r_3-J^r_4)\right],
%\nonumber \\
\label{self02T06}
\eea
where
\bea
   J_1^r&\equiv&\int\frac{k^4dk}{(2\pi)^2}\frac{n_b(E_1)}{2E_1}r(E_1-rp_0)
\nonumber \\
   &&\hspace{-1cm}\times
   \left\{
     \frac{-2rE_1p_0+p_0^2+p^2+m_y^2-m_{\psi_y}^2}{2k^2p^2}
\right.
\nonumber \\
&&\hspace{-0.7cm}
  +\left[1-\frac{(-2rE_1p_0+p_0^2+p^2+m_y^2-m_{\psi_y}^2)^2-4k^2p^2}{4k^2p^2}
\right]
\nonumber \\
&&\hspace{-0.7cm}\times\Theta(rE_1,k,1)\Bigg\},
\eea
\bea
   J_2^r&\equiv&\int\frac{k^4dk}{(2\pi)^2}\frac{n_f(E_2)}{2E_2}r
\nonumber \\
   &&\hspace{-1cm}\times
   \left\{
     \frac{-2rE_2p_0+p_0^2+p^2-m_y^2+m_{\psi_y}^2}{2k^2p^2}
\right.
\nonumber \\
&&\hspace{-0.7cm}
  +\left[1-\frac{(-2rE_1p_0+p_0^2+p^2-m_y^2+m_{\psi_y}^2)^2-4k^2p^2}{4k^2p^2}
\right]
\nonumber \\
    &&\hspace{-0.7cm}\times  \Theta(rE_2,k,-1) \Bigg\},
\eea
\bea
   &&\hspace{-.5cm}J_3^r\equiv\int\frac{k^4dk}{(2\pi)^2}\frac{n_b(E_1)}{2E_1}\frac{1}{p}
\nonumber \\
   &&\hspace{-0.5cm}\times
   \left\{\frac{2}{3}
     -\frac{(-2rE_1p_0+p_0^2+p^2+m_y^2-m_{\psi_y}^2)^2}{3(2kp)^2}
\right.
\nonumber \\
&&\hspace{-.5cm}
  +\left[\frac{(-2rE_1p_0+p_0^2+p^2+m_y^2-m_{\psi_y}^2)^3}{(2kp)^2}
\right.
\nonumber \\
    &&\hspace{-.5cm}-
     (-2rE_1p_0+p_0^2+p^2+m_y^2-m_{\psi_y}^2)
     {\color{white}\frac{1^1 }{1^1}}\hspace{-0.4cm}
    \Bigg]
%\nonumber \\
%    &&\times
\Theta(rE_1,k,1) \Bigg\}
\nonumber \\
\eea
and
\bea
   &&\hspace{-0.1cm}J_4^r\equiv\int\frac{k^4dk}{(2\pi)^2}\frac{n_f(E_2)}{2E_2}\frac{1}{p}
\nonumber \\
   &&\hspace{-.0cm}\times
   \left\{-\frac{2}{3}
     +\frac{(-2rE_2p_0+p_0^2+p^2-m_y^2+m_{\psi_y}^2)^2}{3(2kp)^2}
\right.
\nonumber \\
&&\hspace{-0cm}
  +\left[\frac{(-2rE_2p_0+p_0^2+p^2-m_y^2+m_{\psi_y}^2)^3}{(2kp)^2}
\right.
\nonumber \\
    &&-\left.
     (-2rE_2p_0+p_0^2+p^2-m_y^2+m_{\psi_y}^2)
     {\color{white}\frac{1^1 }{1^1}}\hspace{-0.4cm}
    \right]
%\nonumber \\
%    &&\times \left.
\Theta(rE_2,k,-1)
%   {\color{white}\frac{1^1 }{1^1}}\hspace{-0.4cm}\right
   \Bigg\}.
\nonumber \\
\eea
In the HTL approximation \Eq{self02T06} becomes
\bea
(\Sigma_T^2)_J&=&4h^2
      \left\{i\gamma_4 \mathcal{T}_4-\gamma_i\hat{p}_i\mathcal{T}_2
      \right\}\frac{\pi T}{m_y^3},
\label{self02T07}
\eea
with
\bea
\mathcal{T}_4&\equiv&
      \frac{-4p^3p_0+10pp_0^3}{32p^3(p^2-p_0^2)}
\nonumber \\
     &&+\frac{(p^2-p_0^2)(5p^2+4p_0^2)Q_0\left(\frac{p_0}{p}\right)}{32p^3(p^2-p_0^2)}
%\nonumber \\
\eea
and
\bea
\mathcal{T}_2&\equiv&
    \frac{-14p^5+9p^3p_0^2+14pp_0^4}{48p^3(p^2-p_0^2)}
\nonumber \\
      &&+\frac{3(7p^4p_0-2p^2p_0^3-5p_0^5)Q_0\left(\frac{p_0}{p}\right)}{48p^3(p^2-p_0^2)},
\eea
where we used the identity
\bea
   \int_0^\infty \frac{dk}{E^\alpha}\frac{n_b(E)}{E}\simeq
   \frac{1}{2}\mu^{\epsilon}\pi^{(1-\epsilon)/2}Tm^{-\alpha-1-\epsilon}
    \frac{\Gamma\left(\frac{\alpha+1+\epsilon}{2}\right)}{\Gamma\left(\frac{\alpha+2}{2}\right)}.
\nonumber \\
\eea
The last result completes the calculation of the fermion
self-energy with thermal and magnetic effects. In appendix B, after we diagonalize the heavy fermion propagator with thermal and magnetic effects, we shall obtain explicitly the thermo-magnetic mass of the heavy fermion.

The boson self-energy is given by
\bea
   m_b^2(T,B)=\Pi_a+\Pi_b,
\label{selfboall0}
\eea
where
\bea
   \Pi_a=-h^2\sumint\frac{d^4K}{(2\pi)^4}Tr[\gamma_5S_B(K)\gamma_5S_B(K-P)]
\nonumber\\
\eea
and
\bea
   \Pi_b= 4h^2\sumint\frac{d^4K}{(2\pi)^4}\Delta_B(K).
\eea
In the  HTL approximation each contribution in  \Eq{selfboall0} reduces to
\bea
   \Pi_a&=&-4h^2\left(-\frac{T^2}{24}
           -\frac{7\zeta(3)}{12(2\pi)^4}\frac{(eB)^2}{T^2}\right)
\label{pia}
\eea
and
\bea
   \Pi_b&=&4h^2\left(\frac{T^2}{12}-\frac{m_yT}{4\pi}
      -\frac{m_y^2}{(4\pi)^2}
      \left[\ln\left(\frac{m_y^2}{(4\pi T)^2}\right)
     \right.\right.
\nonumber \\
        &&\left.+2\gamma_E-1\Bigg]
        -\frac{T(eB)^2}{96\pi m_y^3}\right).
\eea

In this way, we have for the boson self-energy:
\bea
   m_b^2(T,B)&=&4h^2\left(\frac{T^2}{8}-\frac{m_y T}{4\pi}
     -\frac{m_y^2}{(4\pi)^2}
      \left[\ln\left(\frac{m_y^2}{(4\pi T)^2}\right)
   \right.\right.
   \nonumber \\
         &&\left.+2\gamma_E-1\Bigg]
         -\frac{T(eB)^2}{96\pi m_y^3}
         \right),
\label{selfboall}
\eea
where we have dropped the last subleading term in \Eq{pia}.

\section{Thermo-magnetic effective potential}

The one-loop effective potential is obtained by adding up the fermion and
boson contributions
\bea
   V_{eff}&=&V_b+V_f,
\label{vtotal}
\eea
where
\bea
   V_b=\int\frac{d^4p}{(2\pi)^4}\ln\left[p^2-m_\chi^2+m_b^2(T,B)\right]
\eea
and
\bea
   V_f=-\frac{1}{2}\int\frac{d^4p}{(2\pi)^4}\ln
   Det\left[\slsh{p}-m_{\Psi_\chi}+\Sigma^{B}(p)\right],
\label{vferfull}
\eea
with $\Sigma^B(p)$ and $m_b^2(T,B)$ given by
Eqs.~(\ref{selfall}) and~(\ref{selfboall}), respectively. The fermion determinant has the form
\bea
  &&\hspace{-1cm}Det\left[\sum_{r=\pm1}{\not \! A}^r\Delta(r)-m_{\Psi_\chi}\right]
\nonumber \\
   &=&m^4+({A_0^+}^2-{A_3^+}^2)({A_0^-}^2-{A_3^-}^2)
\nonumber \\
    &&+(A_1^++iA_2^+)^2(A_1^--iA_2^-)^2
\nonumber \\
    &&-2(A_0^+A_0^- - A_3^+A_3^-)(A_1^+ + iA_2^+)(A_1^- - iA_2^-)
\nonumber \\
    &&-m^2\left[
         {A_0^+}^2+{A_0^-}^2 - {A_3^+}^2-{A_3^-}^2
\right.
\nonumber \\
    &&\hspace{1cm}\left.-2(A_1^+ + iA_2^+)(A_1^- - iA_2^-)^{ }
          \right],
\eea
where $\Delta(r)\equiv(1+ir\gamma^1\gamma^2)/2$ and
\bea
   A_0^r&=&p_0
        -\frac{m_f^2}{2p}
         \left[Q_0+\frac{r eB}{\pi m_y T}\frac{2p^3p_0}{(p_0^2-p^2)^2}
\right.
\nonumber \\
&&\left.
       +\frac{(eB)^2}{4\pi T m_y^3}\frac{1}{p^2}
        \left((p^2+2p_0^2)Q_0
       +\frac{5p p_0^3-2p^3 p_0}{p^2-p_0^2}
        \right)
        \right],
\nonumber \\
\eea

\bea
   A_3^r&=&p_3
        +\hat{p}_3\frac{m_f^2}{2p}
         \left[1-\frac{p_0}{p}Q_0-\frac{r eB}{\pi m_y T}\frac{p^2(p^2+p_0^2)}{(p_0^2-p^2)^2}
\right.
\nonumber \\
&&\hspace{-1cm}\left.
       -\frac{(eB)^2}{4\pi T m_y^3}\frac{p_0}{2p^3}
        \left((11p^2+10p_0^2)Q_0
       -\frac{10p^5-28p p_0^4}{3p_0(p^2-p_0^2)}
        \right)
        \right],
\nonumber \\
\eea

\bea
   A_{1,2}^r&=&p_{1,2}
        +\hat{p}_{1,2}\frac{m_f^2}{2p}
         \left[1-\frac{p_0}{p}Q_0
       -\frac{(eB)^2}{4\pi T m_y^3}\frac{p_0}{p^3}
\right.
\nonumber \\
&&\hspace{-0.5cm}\left.
        \times\left((7p^2+5p_0^2)Q_0
       -\frac{14p^5-9p^3 p_0-14p p_0^4}{3p_0(p^2-p_0^2)}
        \right)
        \right].
\nonumber \\
\eea
In the infrared limit this determinant reduces to
\bea
  &&\hspace{-1cm}
   Det\left[\sum_{r=\pm1}{\not \! A}^r\Delta(r)-m_{\Psi_\chi}\right]
\nonumber \\
    &&\hspace{2.5cm}\simeq \prod_{r=\pm1}\left[p_\mu p^\mu-m_{\Psi_\chi}^2(T,B)\right],
\nonumber \\
\label{detinfra}
\eea
where
%\bea
%   m_{\Psi_\chi}^2(T,B;r) &\equiv& m_{\Psi_\chi}^2
%\nonumber \\
%+
%     \frac{h^2T^2}{2}
%m_f^2
%\left(1-\frac{r}{3}\frac{eB}{\pi m_y T}
%       +\frac{11}{12}\frac{(eB)^2}{\pi m_y^3 T}
%     \right)
%\nonumber \\
%\eea

\bea
   m_{\Psi_\chi}^2(T,B;r) &\equiv& m_{\Psi_\chi}^2
+m_f^2(T,B;r),
\eea
with
\bea
m_f^2(T,B;r)\equiv m_f^2
\left(1-\frac{r}{3}\frac{eB}{\pi m_y T}
       +\frac{11}{12}\frac{(eB)^2}{\pi m_y^3 T}
     \right)
\nonumber \\
\eea
accounts for thermal and magnetic effects to the heavy fermion mass.
Replacing \Eq{detinfra} into \Eq{vferfull}, the
integration over all momenta in Euclidean space in \Eq{vtotal} becomes
straightforward and we obtain
\bea
   V_\chi(\phi,T,B)&=& \frac{1}{32\pi^2}\Bigg[m_\chi^4(T,B)\ln\left(\frac{m_\chi^2(T,B)}{\Lambda_{uv}^2}\right)
\nonumber \\
&&\hspace{-1.5cm}
  -\frac{1}{2}\sum_{r=\pm 1}
    m_{\Psi_\chi}^4(T,B;r)\ln\left(\frac{m_{\Psi_\chi}^2(T,B;r)}{\Lambda_{uv}^2}\right)\Bigg]
   + C,
\nonumber \\
%&&+const.
\label{eq:potential}
\eea
where we have introduced the ultraviolet cutoff $\Lambda_{uv}$ that
together with the constant $C$ are determined from the renormalization
conditions discussed below. Note that the main divergences cancel out
and the remaining ones are due to the soft SUSY breaking term which we
have defined as the slight difference between the fermion and boson
masses, that is
\be
m_\chi^2= m_{\Psi_\chi}^2 + M_s^2,
\label{eq:susybreakingap}
\ee
where $M_s^2$ is a small contribution compared with $m_\chi^2$.

From Eq.~(\ref{eq:potential}) one can note that if
supersymmetry is not broken, the effective potential vanishes. This happens since in this case
the fermion and boson masses, $m_{\psi_\chi}$ and $m_\chi$, are the
same, and in the HTL approximation both self-energies,
$m_f(T,0)$ and $m_b(T,0)$ are equal.

By imposing for $T=0$ and $eB=0$ the conditions
\be
\left.V_\chi (\phi,T,B)\right|_{\phi=\phi_0} =0,
\ee
\be
\left.\frac{\partial }{\partial \phi} V_\chi (\phi,T,B)\right|_{\phi=\phi_0}=0,
\ee
to the potential in Eq.~(\ref{eq:potential}), and neglecting terms
proportional to $M^4_s$ and $(m_f^2(T,B)- m_b^2(T,B))^2$, we finally get
\bea
&&\hspace{-1cm}V_\chi(\phi, T, B)=\frac{M_s^2\left(m_{\Psi _\chi}^2 + m_b^2(T,B)\right)}{16\pi^2}
\nonumber \\
&&\hspace{-.5cm}\times\left(1-\sum_{r=\pm1}\frac{m_f^2(T,B;r)-m_b^2(T,B)}{2M_s^2}\right)
\nonumber \\
&&\hspace{-0.5cm}\times\left(\ln\left(\frac{m_{\Psi_\chi}^2+m_b^2(T,B)}{{m_0}_{\Psi_\chi}^2}\right)-1\right)+\frac{M_s^2{m_0}_{\Psi_\chi}^2
}{16\pi^2}.
\nonumber \\
\eea
We have defined ${m_0}_{\Psi_\chi}\equiv \sqrt{2}g\phi_0$.

\end{document}